# The Effect of Frame Rate on 3D Video Quality and Bitrate

Amin Banitalebi-Dehkordi, Mahsa T. Pourazad, and Panos Nasiopoulos

**Abstract** — Increasing the frame rate of a 3D video generally results in improved Quality of Experience (QoE). However, higher frame rates involve a higher degree of complexity in capturing, transmission, storage, and display. The question that arises here is what frame rate guarantees high viewing quality of experience given the existing/required 3D devices and technologies (3D cameras, 3D TVs, compression, transmission bandwidth, and storage capacity). This question has already been addressed for the case of 2D video, but not for 3D. The objective of this paper is to study the relationship between 3D quality and bitrate at different frame rates. Our performance evaluations show that increasing the frame rate of 3D videos beyond 60 fps may not be visually distinguishable. In addition, our experiments show that when the available bandwidth is reduced, the highest possible 3D quality of experience can be achieved by adjusting (decreasing) the frame rate instead of increasing the compression ratio. The results of our study are of particular interest to network providers for rate adaptation in variable bitrate channels.[1]

## 1. Introduction

3D video technologies have been welcomed by the consumer market in the recent years. As these technologies mature, the consumers' appetite for high quality 3D viewing experience is elevating. In order to achieve the best possible results, the factors that attribute to 3D quality need to be carefully taken into consideration throughout the 3D content delivery pipeline (capturing, transmission, and display). There are several factors and parameters that affect the perceptual quality of 3D content. Some of these factors are explicit to 3D content and do not affect the quality of 2D content or do not exist in the case of 2D, such as disparity, display size, 3D display technology (active, passive, glasses-free), and binocular properties of Human Visual System (HVS). Other factors are common attributes between 2D and 3D recording, but may have different effect on these two types of media, such as brightness [1] and color saturation [2]. Although fast moving objects may have a negative impact in the visual quality of 2D content, this quality degradation effect due to fast motion seems to be significantly magnified in the case of 3D [3,4]. The reason is that fast motion in 3D results in rapid change in the perceived depth of objects, if the motion direction is perpendicular to the screen. This in turn leads to fast decoupling of vergence and accommodation (which is the main source of visual fatigue when watching 3D), resulting in degradation of the overall 3D quality of experience [3,4].

Considering that motion of objects cannot be controlled in live 3D videos and in the case of movies, motion might be one of the key elements of the story line, the need for new tools that make motion of objects appear smoother and improve the 3D viewing Quality of Experience (QoE) has become noticed by the research community. To address this need, the film industry has introduced higher frame rates for 3D video capturing recently [5]. A stereoscopic video captured with higher frame rate than the traditional one, i.e., 24 frames per second (fps), is likely to appear sharper, less blurred, more natural, and making the viewing experience less cumbersome and more comfortable [6]. However, existing studies and subjective evaluations on 2D videos show that the human eye is not able to distinguish the 2D videos with standard conventional frame rates (25 fps (PAL), 30 fps (NTSC), and 24 fps (cinema films)) from the ones with higher frame rates, or the difference in perceptual quality is not significant [7-12]. For this reason, sacrificing bandwidth or memory storage for supporting higher frame rates 2D video content has not been suggested.

While the effect of frame rate on 2D perceptual video quality has been studied for many years [7-12], in the case of 3D there are still unanswered questions. The main question is identifying what frame rate yields the best visual quality for 3D. Recent feedback on 3D movies captured at 48 fps indicates improved 3D viewing experience [6,13], but the question that remains is how much improvement is achieved by increasing the frame rate from 24 fps to 48 fps, or if any frame rate increase above 48 fps will result in additional visual quality improvement. For the case of 3D content broadcasting, in addition to the questions regarding the impact of frame-rate increase on 3D video quality, there are questions on bandwidth requirements for the transmission of high frame rate 3D content. Considering that the required bandwidth for the transmission of 3D video is generally higher than that of 2D video, it is important to perform feasibility studies on high frame rate 3D content transmission and come up with bitrate adaptation guidelines for variable bitrate networks. These guidelines, similar to the existing ones for the transmission of 2D video [7-12,14-16], will help with adjusting the content frame-rate adaptively (dropping frame rate or frame rate up-conversion), according to channel capacity and required video quality. Limited work has been done in this regard for 3D video and, therefore, there is still room for improvement [17,18].

To address some of the questions on the effect of frame rate on 3D video quality, in our previous study we compared the perceptual quality of 3D videos with conventional frame rates of 24 fps, 30 fps, and 48 fps with that of 60 fps through subjective tests [13]. Our objective was to investigate how much the viewing quality of experience is improved by switching from 24 fps, 30 fps, or 48 fps to 60fps, and determine if there is a significant difference. In our study, the frame rates of 24 fps and 30 fps are chosen particularly because these frame rates have already being used in the 3D industry, and the frame rates of 48 fps and 60 fps were part of our experiment as there is a growing interest towards capturing 3D content with such frame rates. In the conducted experiments, the subjects were asked for each scene to compare the quality of captured 3D videos at 60 fps with that of videos captured at lower frame rates (24, 30, or 48 fps). In other words, this experiment was designed so that the 60 fps 3D videos were used as the reference and the quality of the 3D videos from the same scenes, but with lower frame rate were ranked with respect to them. The experiment results showed that there is no statistical difference between the Mean Opinion Score (MOS) values of 48 fps 3D videos and 60 fps ones of the same scene, however the MOS values of these frame rates are much higher than those of 24 and 30 fps 3D videos. These findings confirm that subjects prefer 60 fps 3D videos over 24 and 30 fps 3D videos.

In this paper, we further investigate the effect of the frame rate on 3D viewing experience with the objective to identify the appropriate frame rate for 3D video capturing. To this end, we analyze the relationship between frame rate, bitrate, and the 3D QoE through extensive subjective tests. The findings of our study are helpful in defining bitrate adaptation guidelines for 3D video delivery over variable bitrate networks. These guidelines will allow network providers to change 3D content frame rate in order to deal with bandwidth capacity changes so that viewers' quality of experience is not significantly affected.

The rest of this paper is organized as follows: Section II explains the procedure to prepare the 3D video test set, Section III provides details on the experiment procedure, Section IV contains the results and discussion, and Section V concludes the paper.

## 2. Preparation of the 3D Video Data Set

This section provides details on the capturing and preparation of the video data set used in this study, including hardware configurations as well as post-processing steps.

### 2.1 Camera Configuration

In order to capture 3D videos for our experiments, we use four cameras of a same model, with identical firmware and camera settings. The cameras are mounted on a custom-made bar and are aligned in parallel. One camera pair is configured to capture 60 fps (two side-by-side cameras on the right side of the bar in Fig. 1) and the other camera pair is set to capture 48 fps (two side-by-side cameras on the left side of the bar in Fig. 1). To generate 30 fps and 24 fps stereoscopic videos, the captured 60 fps and 48 fps stereoscopic videos are then temporally down-sampled by a factor of two. This is done by starting from the first frame and dropping every other frame in each video. As a result, we obtain 3D videos at four different frame rates of 24, 30, 48, and 60 fps from the same scene. These frame rates are chosen, because they are available options in consumer cameras. Presently, theater content is shot in 24 or 30 frames per second while there is interest to know the effect of 48 fps and 60 fps for 3D.

### 2.2 Database Capturing

In our study, GoPro cameras are chosen for capturing the test dataset, because of their small size (which allows us to minimize the difference between the captured stereo pairs) and their capability of capturing high-resolution (HD) videos (1080x1920) at up to 60 frames per second (fps). Since the camera lenses are almost identical and have the same f-number, the camera shutter speed (exposure time) controls the amount of light that reaches the sensor. The shutter speed in these cameras is automatically set to the inverse of the video frame rate [19]. GoPro cameras come with a built-in wide-angle lens, which may cause a fisheye effect at the borders of the picture. During capturing, special attention was given to the contextually important areas to ensure they were not affected by fisheye distortions. This was further enforced by applying the 3D Visual Attention Model (3D VAM) described in [20] to identify the visually important areas of the captured videos. Videos whose visually important areas are affected by fisheye distortion, were excluded from our database. Considering that identical cameras are used for capturing the test dataset, the same amount of fisheye effect exists in all the different frame-rate versions for the same scene. This allows us to conduct a fair comparison among different frame-rate versions of the same scene and studying the effect of frame rate on 3D visual perception.

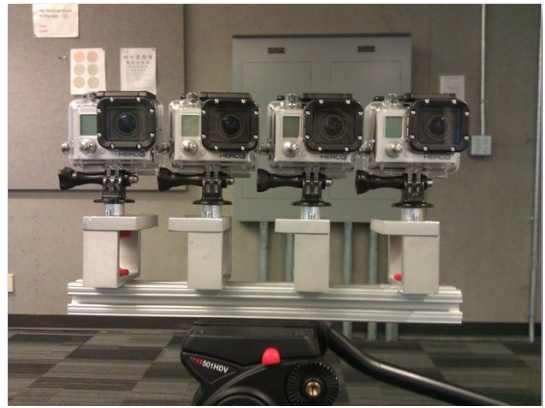

**Fig. 1.** Camera configuration

At the time of capturing, it is ensured that there is no window violation (when part of an object is popping out of the screen, which causes the brain to get confused because of two contradictory depth cues) by properly selecting the framing window.

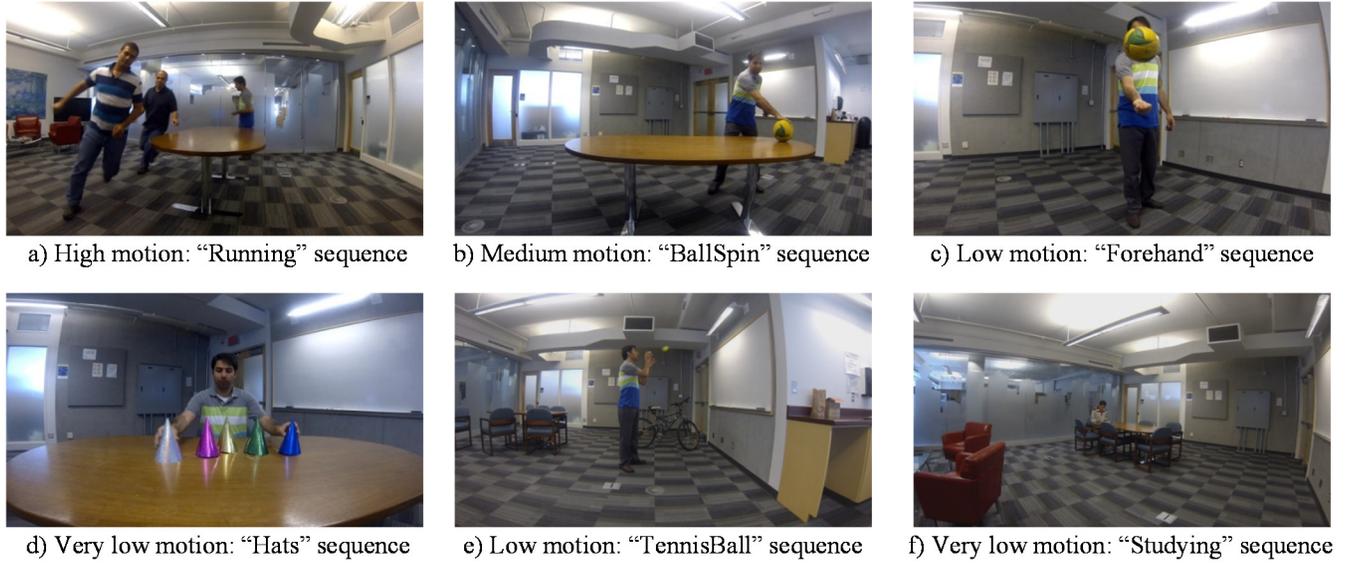

**Fig. 2.** Snapshots of the left views of the 3D video database

Six indoor scenes are captured using the camera setup shown in Fig. 1, each 10 seconds long. The resolution of the original 3D videos is 1920×1080 (High Definition) for each view and the baseline between cameras are set at 7 cm. A snapshot of the left view of each scene is shown in Fig. 2. Our database is publicly available at [21]. TABLE 1 provides specifications about the captured videos. For each video sequence, the amount of spatial and temporal perceptual information is measured according to the ITU Recommendation P.910 [22] and results are reported in TABLE 1. For the spatial perceptual information (SI), first the edges of each video frame (luminance plane) are detected using the Sobel filter [23]. Then, the standard deviation over pixels in each Sobel-filtered frame is computed and the maximum value over all the frames is chosen to represent the spatial information content of the scene. The temporal perceptual information (TI) is based upon the motion difference between consecutive frames. To measure the TI, first the difference between the pixel values (of the luminance plane) at the same coordinates in consecutive frames is calculated. Then, the standard deviation over pixels in each frame is computed and the maximum value over all the frames is set as the measure of TI. More motion in adjacent frames will result in higher values of TI. Note that the reported values for spatial and temporal information measures are obtained from the 60 fps version of each sequence, as this version is closer to our visual true-life perception. Fig. 3 shows the spatial and temporal information indexes of each test sequence, as indicated in [22].

For each sequence shown in TABLE 1, we also provide information about the scene's depth bracket. The depth bracket of each scene is defined as the amount of 3D space used in a shot or a sequence (i.e., a rough estimate of the difference between the distance of the closest and the farthest visually important objects from the camera in each scene) [24].

**TABLE 1**
DESCRIPTION OF THE 3D VIDEO DATABASE

| Sequence | Resolution | Spatial Complexity (Spatial Information [22]) | Temporal Complexity (Temporal Information [22]) | Depth Bracket | Motion Level |
|---|---|---|---|---|---|
| Running | 1920×1080 | High (49.22) | High (22.19) | Wide | High |
| BallSpin | 1920×1080 | High (44.35) | Medium (12.05) | Medium | Medium |
| Forehand | 1920×1080 | Medium (34.39) | Low (5.89) | Medium | Low |
| Hats | 1920×1080 | Medium (35.93) | Medium (10.15) | Narrow | Very low |
| TennisBall | 1920×1080 | High (44.27) | Very low (3.45) | Wide | Low |
| Studying | 1920×1080 | High (44.17) | Very low (2.87) | Wide | Very low |

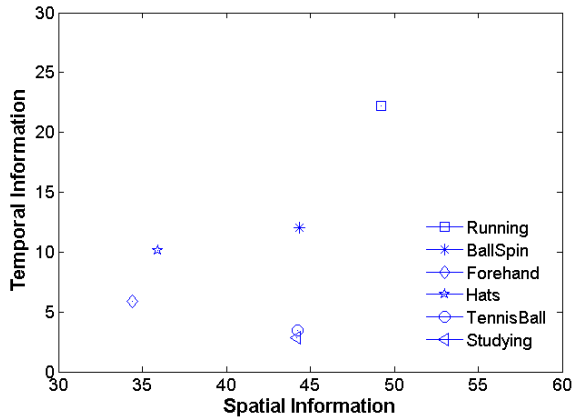

**Fig. 3.** Distribution of the spatial and temporal information over the video database

The captured 3D streams are post-processed to ensure that they are temporally synchronized, rectified, and comfortable to watch. The following subsections elaborate on the applied post-processing schemes.

### 2.3 Temporal Synchronization

To temporally synchronize the cameras, a single remote is used to control the four cameras together so that they all start and finish recording at the same time. However, in practice there are cases where, due to lack of timing accuracy between the remote and cameras, the captured videos are not completely temporally synchronized. In these cases, manual correction is applied to remove a few frames from the videos and achieve temporal synchronization. Considering that the videos are originally captured at 48 fps and 60 fps, manual correction achieves visually acceptable temporal synchronization.

Note that temporal synchronization is performed before we temporally down-sample the captured videos to 24 fps and 30 fps.

### 2.4 Alignment of the 3D Content

Vertical parallax in stereoscopic video makes viewers uncomfortable, as fusing two views with vertical parallax is difficult for the brain. To reduce the vertical parallax, the four cameras are physically aligned by using identical screws to mount them on a horizontal bar (see Fig. 1). This reduces the vertical parallax to some extent, but the videos may still suffer from some vertical misalignment.

To remove the vertical parallax, the left and right views are rectified using an in-house developed software solution. Our approach first extracts the features of the first frame of the left and right views using the Scale Invariant Feature Transform (SIFT) [25]. The features of the left frame are matched to the features of the right frame. The top 10% of all matching features, whose vertical disparities are considerably different from the median disparity value of all matching features, are detected as outliers. These outlier features are removed to ensure the stability of the algorithm. The Cartesian coordinates of rest of the matching features are saved. The median of all the *y* coordinates of matching points between the two frames, *dy,* is the amount of pixels that each original frame need to be shifted vertically. More specifically, the median vertical mismatch of the matching points gives an estimate of how much each of the views needs to be cropped so that the resulting cropped images contain rectified views without vertical parallax.

Note that since the cameras used for capturing have identical fixed focal length and no digital zoom function, the recorded views do not need zoom correction.

### 2.5 Disparity Correction

When 3D videos are captured by parallel stereo cameras, all the objects pop out of the screen as the cameras converge at infinity. In this case, the captured objects are known to have a negative horizontal parallax. This negative parallax occurs when the left-view of an object is located further to the right than the right-view version of the same object. Existing studies show that when objects appear to be in front of the screen for a considerable amount of time they induce visual discomfort [24].

It is a good practice to modify the disparity information (disparity correction) of the content in order to relocate the 3D effect behind the display [24]. To this end, the left frames need to be shifted towards the left and the right frames towards the right, so that the negative horizontal parallax of 3D videos is reduced [24]. To avoid black lines on the vertical edges of the frames, the content is cropped to match the aspect ratio and then it is scaled up. To determine the amount of pixels by which each original frame will be shifted horizontally (i.e., $dx$), we find the largest negative value of all the $x$ coordinates of matching points between the two frames [24]. The negative number with the largest absolute value of the $x$ coordinates represents the photographed point in space that is closest to the cameras ($d_{min}$). Once the frames are shifted according to $dx$, they are cropped and then enlarged using bicubic interpolation so that they maintain their original size before the shifting (1080×1920 pixels) [24]. Considering that $d_{min}$ changes over frames in some of the scenes, the shifting parameter ($dx$) is determined based on the frame with the smallest $d_{min}$ and then the same amount of cropping is applied to the rest of the frames. This disparity correction process can improve the 3D quality of experience by 19.86% on average [24]. The effect of disparity correction is mainly reflected in reducing the 3D visual discomfort, which is caused when the eyes try to focus on the screen (accommodation), while the eyeballs try to converge on objects (vergence) that are popping out of the screen. In other words, disparity correction may only shift objects along the depth direction to push them inside the comfort zone.

## 3. Subjective Experiments Procedure

The effect of the frame rate on 3D QoE and bitrate was studied through two series of subjective tests using the captured 3D video database. The following subsections elaborate on our experiments.

### 3.1 Case Study I: Effect of the Frame Rate on the Quality of 3D Videos

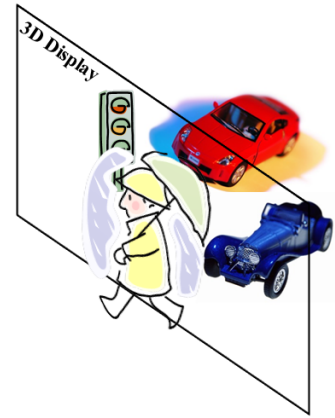

viewer

**Fig. 4.** Disparity correction mechanism: Objects are pushed to the 3D viewing comfort zone

In the first experiment, the goal is to study the relationship between the frame rate and quality of 3D videos and identify the appropriate frame rate for 3D video capturing. To this end, subjective tests are performed to evaluate the visual quality of the 3D test videos at different frame rates. Here, for each scene, the 3D videos with different frame rates are shown one by one and subjects are asked to rate the 3D quality of each video separately and independently from the other videos. To determine if there is a preferred frame rate for 3D viewing, no reference high frame-rate video is provided (unlike [13]). For this case study, five 3D scenes are selected from the captured database and each scene is shown at all four different frame rates (total of 20 stereoscopic videos). The selected scenes are "TennisBall", "BallSpin", "Forehand", "Running", and "Studying".

### 3.2 Case Study II: Effect of the Frame Rate on the Quality of the Compressed 3D Videos

In the second experiment, the objective is to study the effect of frame rate at different compression levels on the 3D quality of experience. This study allows determining at each bitrate level what frame rate results in the highest 3D quality of experience. To this end, the video scenes captured at different frame rates are compressed at a variety of bitrates, and subjective tests are performed to evaluate the quality of the compressed 3D videos.

The 3D video sequences are encoded using the emerging 3D HEVC standard (3D-HTM 8.0 reference software [26]) [27,28]. The Quantization Parameter (QP) is set according to the suggested Common Test Conditions by JCT-3V (a joint group under MPEG (ISO/IEC Moving Picture Experts Group) and VCEG (Video Coding Experts Group)) to four different levels of 25, 30, 35, and 40 [29]. The random access high efficiency configuration is used, while the GOP (Group Of Pictures) size is set to eight. Moreover, ALF (Adaptive Loop Filter), SAO (Sample Adaptive Offset), and RDOQ (Rate-Distortion Optimized Quantization) are enabled [27,28]. In addition, in order to have a fair comparison, the encoding parameters are adjusted according to the frame rate of the 3D videos. For instance, the "intra period" parameter (number of P-frames or B-frames between every two consecutive I-frame) for 24 fps, 30 fps, 48 fps, and 60 fps videos is set at 24, 32, 48, and 64, respectively, to ensure that the size of the intra period is proportional to the frame rate and at the same time is a multiple of the GOP size (i.e., 8). In this case study, four 3D scenes are chosen from the database and for each scene, all the 3D videos with different frame rates (four frame rates) are compressed at four QP levels. As a result, the test set includes a total of 64 3D videos. The selected scenes are "Running", "BallSpin", "Forehand", and "Studying".

### 3.3 Test Procedure

Both experiments were conducted according to the viewing conditions specified by the ITU-R recommendation BT.500-13 [30]. Sixteen subjects participated in the first experiment and another eighteen in the second one. The subjects' age ranged from 19 to 29 years old. Before the experiments, all subjects were screened for visual acuity (using Snellen chart), color blindness (using Ishihara chart), and stereovision acuity (using Randot test) and passed the required thresholds. The 3D display used for the experiments was a 64" full HD (High Definition) 3D TV with circularly passive polarized glasses. The screen resolution is the same as the resolution of the videos (1080x1920, which corresponds to an area of 168.2x87.5 cm on the screen) and therefore there was no need for scaling the videos.

Test sessions were based on the Single Stimulus (SS) method, in which subjects view videos of the same scene with different frame rates in random order. Note that in both case studies, each test session included one randomly selected test video from all the scenes. Thus, the chances that subjects could become biased or exhausted watching the same scene are reduced, while the test sequences are randomized.

Grading was performed according to the Numerical Categorical Judgment (NCJ) method [30], where observers rate video quality based on a discrete range from 0 to 10 (0 representing the lowest quality and 10 representing the highest quality) [30]. As suggested by Quan et al. [31], it was explained to the subjects that the term "quality" in general means how pleasant they think a video looks. Specifically, they were asked to rate the quality based on a combination of different factors such as "naturalness" [31,32], "comfort" [33], "depth impression", "sharpness", and "temporal smoothness" [24,31,34]. There was a "training" session before the "test" session, so that the subjects become familiar with the videos and the test structure. During the training period participants were explained how/what to grade watching each test video. In order to minimize the effect fisheye distortions could have on the subjective evaluations, information about the fisheye effect was given to the subjects to familiarize them with this type of distortion and thus help them judge the perceptual quality of the videos without taking into account the fisheye effect. Following what is considered common practice in such tests, even though a training session was provided before each test session, a few "dummy" sequences were shown at the beginning of each test session [30]. The scores for the dummy sequences were excluded from the analysis, as their objective is to familiarize the subjects with the test procedure at the beginning of the test session.

After collecting the subjective test results, the outlier subjects were detected and their scores were removed from the analysis. Outlier detection was performed according to the ITU-R BT.500-13 recommendation, Annex 2 [30]. In the outlier detection process, the kurtosis coefficient is calculated to measure how well the distribution of the subjective scores can be represented using a normal distribution. Through this process it was found that there was no outlier in the first experiment, while there were two outliers in the second case study.

## 4. Results and Analysis

Once the experiment data is collected, the Mean Opinion Score (MOS) for each video is calculated as the average of the scores over the subjects set. In order to ensure the reliability of these measurements, a confidence interval of 95% is calculated [30].

### 4.1 Case I: 3D Quality versus Frame Rate for Uncompressed 3D Sequences

In the first case study, the quality of the original video set with different frame rates was subjectively evaluated. Fig. 5 shows the average perceived 3D quality at different video frame rates for the entire video database with 95% confidence interval. As it is observed, the 3D videos with frame rates of 48 fps and 60 fps are highly preferred and rated as excellent quality (MOS greater than 8). On the other hand, the 3D videos with the frame rate of 24 fps are rated as poor/fair quality (MOS between 2 and 5). Considering that the MOS of 3D videos at 60 fps with 95% confidence interval can reach 9.8, one could conclude that increasing the frames rates of 3D videos more than 60 fps may not result in visually distinguishable quality for viewers and will just increase the complexity of capturing, transmission, and display. It is also observed that there is a significant difference between the quality of 3D videos with 24 fps and the ones with 48 fps and 60 fps. In particular, average MOS-difference between videos in 60 fps and videos in 24 fps is around 5.8 (out of 10), indicating a high preference for these high rates.

In order to understand the effect of motion on the perceived 3D video quality, we plot the 3D quality versus frame rate for two videos with low and high levels of motion (sequence "Running" for fast motion and sequence "Studying" for slow motion) in Fig. 6. It can be seen that the difference between the MOS values at 30 fps and 48 fps increases when the motion level is high. In particular, 3D quality drops by 3.9 in terms of MOS when the frame rate decreases from 48 fps to 30 fps for the video with fast motion, whereas the quality drops only by 1.5 in terms of MOS in the case of low-motion video.

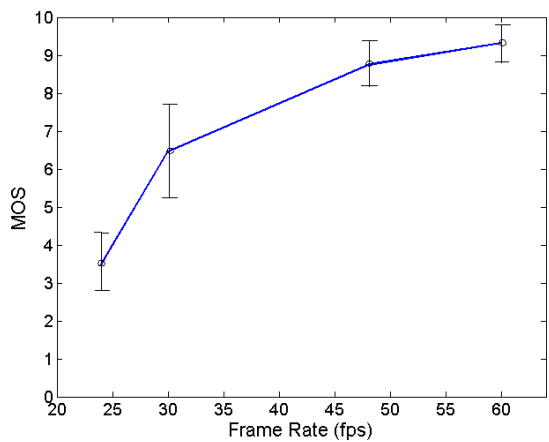

**Fig. 5.** Average perceived 3D video quality (MOS) at different frame rates

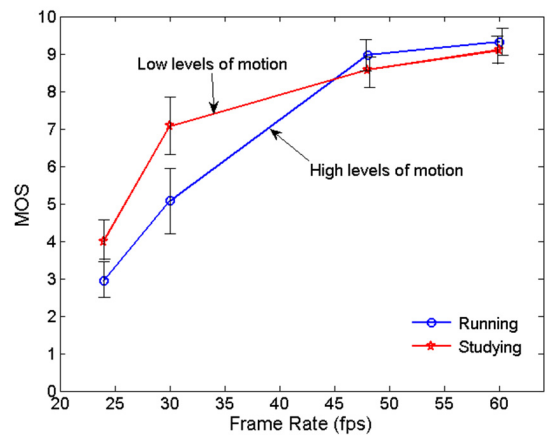

**Fig. 6.** 3D video quality (MOS) at different frame rates for videos with low ("Studying" sequence) and high ("Running" sequence) levels of temporal motion

In other words, when a scene contains fast moving objects, low frame rates (in this case 24 fps and 30 fps) result in an unpleasant 3D experience. Based on this observation it is recommended to capture 3D scenes with high motion at higher frame rates than 30fps to ensure the motion in the scene appears smooth and the 3D quality of experience is improved.

## 4.2 Case II: 3D Quality versus Frame Rate and Bitrate for Encoded 3D Sequences

In the second case study, the quality of compressed 3D videos with different frame rates is subjectively evaluated. After collecting the results and removing the outliers, the average MOS for each video is calculated at different frame rates and bitrates. Fig. 7 illustrates the relationship between 3D quality of experience and frame rate at different bitrates for 3D video sequences with variety of motion levels. By comparing the results for different video sequences, it is observed that in general 3D videos with higher bitrates and higher frame rates are more pleasant to viewers. Another useful observation derived from Fig. 7 is that, except for very low

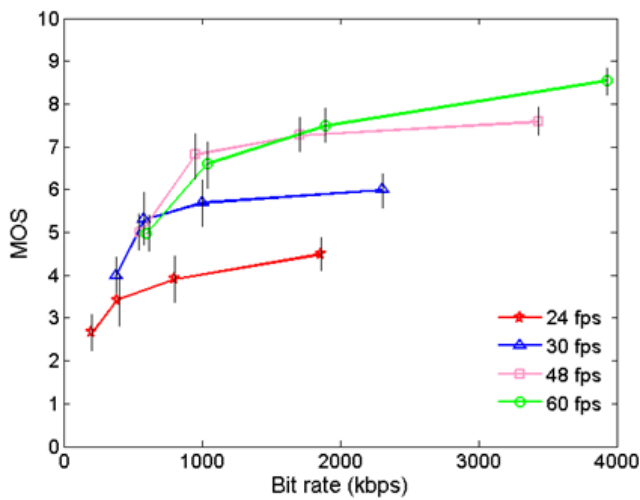
a) High motion: "Running" sequence

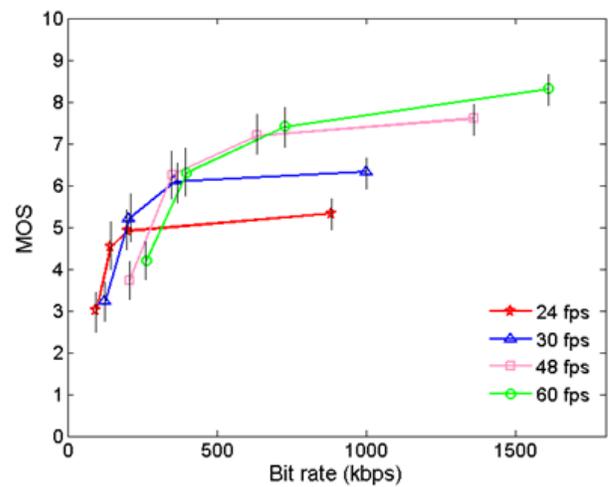
b) Medium motion: "BallSpin" sequence

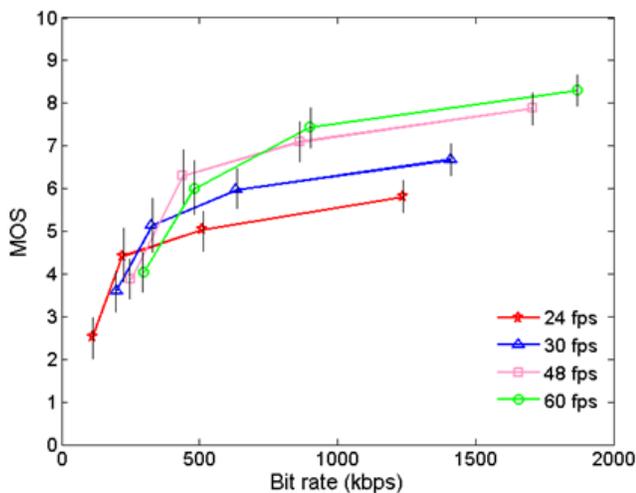
c) Low motion: "Forehand" sequence

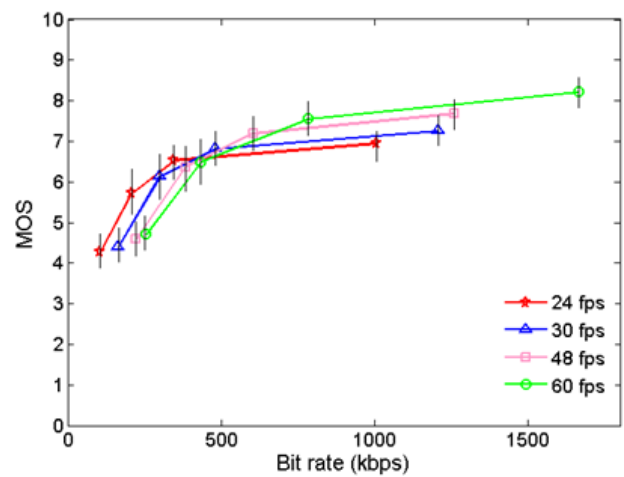
d) Very low motion: "Studying" sequence

**Fig. 7.** 3D quality depicted at different bitrates for various frame rates, in four different levels of motion: a) High motion, b) Medium motion, c) Low motion, and d) Very low motion

bitrates, subjects prefer to watch a high frame rate version of a 3D video rather than its lower frame rate version, even though the high-frame rate one is more compressed.

Moreover, the illustrated results in Fig. 7 for test sequences with different motion levels show that the gap between the perceptual qualities of different frame-rate versions of the same 3D video becomes more significant, if the scene includes higher motion levels. In Fig. 7.a where the test sequence ("Running") includes fast moving objects in the scene, the MOS of the higher frame-rate versions (48 and 60 fps) are higher than those of lower frame rates (24 and 30 fps). The difference becomes quite significant at higher bit rates. Even the highly compressed 60 fps and 48 fps versions of the 3D videos (low bitrate of 1000 kbps) are preferred over the 24 fps version of the same video with slight compression (high bitrate of 2000 kbps or more). This suggests in the case of 3D video content with high motion, to transmit the high frame rate version of the content (if available) at the channel bitrate, instead of the low frame-rate version of the video.

Fig. 7.b and 6.c show the results of our experiment for 3D content with medium and low levels of motion. As it is observed the overall 3D quality of higher frame-rate versions of the sequence is still higher than that of lower frame-rate versions (except at very low bitrates, less than 500 kbps), but the MOS difference is not as high as the case where the motion level of the scene is high. In the case where the motion level in the scene is very low, as it is observed from Fig. 7.d, the perceptual quality of different frame-rate versions of the 3D video sequence are quite similar at the same bitrate level. In other words, frame rate is no longer a contributing factor to the 3D quality and the 3D quality is controlled by bitrate here. This is because when the motion level is low, temporal smoothness provided by frame-rate increase is no longer noticeable.

The subjective test results in Fig. 7 suggest that based on the amount of the available bandwidth (required bitrate), one should choose the appropriate frame rate, which provides the maximum 3D quality. More precisely, to ensure the highest possible 3D quality is achieved at high bitrate, the higher frame rate version of the 3D content shall be transmitted (if available). In case where the bandwidth drops from a very high value, then the frame rate needs to be adjusted (reduced) according to the available bandwidth. At very low bitrates, depending on the application, dropping one of the 3D streams and delivering 2D content is also suggested [35]. Following these guidelines allows network providers to deliver maximum possible 3D video quality by controlling and adjusting the frame rate at different bitrates.

We used the statistical T-test to determine if there is a significant difference between the quality scores obtained from sixteen subjects for different frame-rate versions of each sequence compressed with a specific QP setting. TABLE 2 summarizes the T-test results. The null hypothesis is if the perceptual quality of two different frame rate versions of a test sequence compressed with a specific QP setting is statistically equal. The significance level is set at 0.01. As it is observed for the "Studying" sequence, which has very low motion, there is no strong presumption against the null hypothesis in all QP levels. In other words the perceptual quality of different frame rate versions of "Studying" video sequence compressed with a specific QP setting is statistically equal (with one exception at QP of 25 and frame rate pair of (48, 60)). For the rest of the sequences with different motion levels, the null hypothesis is always rejected for low QPs of 25 and 30 and for frame rate pairs of (24, 30) and (30, 48). This implies that at high bitrates (small QPs), there is a significant difference between the quality of 24fps, 30fps, and 48fps versions of the compressed videos. By comparing the results of 48fps and 60 fps for the test sequences with low to high motion levels, it is observed that the perceptual quality of the 48fps and 60fps versions of the video sequences is equal at all QP levels, except for the case where medium or very high motion level is present and QP is very small (high bitrate). The statistical difference test results also show that in the case of medium, low, and very low motion in a scene, when the bit rate is low (high QP values of 35 and 40), no difference in the subjective quality score of (24, 30) and (30, 48) frame rate pairs is reported (with one exception). In the presence of high motion ("Running" sequence), however, distinct statistical difference is reported when the bit rate is low (QPs of 35 and 40) and frame rate pairs of (24, 30) or (30, 48) are being compared.

**TABLE 2**
STATISTICAL DIFFERENCE (P-VALUES) BETWEEN THE SUBJECTIVE SCORES USING T-TEST

|  | Running (high motion) | | | Ball Spin (medium motion) | | | Forehand (low motion) | | | Studying (very low motion) | | |
| --- | --- | --- | --- | --- | --- | --- | --- | --- | --- | --- | --- | --- |
|  | $P_{60-48}$ | $P_{48-30}$ | $P_{30-24}$ | $P_{60-48}$ | $P_{48-30}$ | $P_{30-24}$ | $P_{60-48}$ | $P_{48-30}$ | $P_{30-24}$ | $P_{60-48}$ | $P_{48-30}$ | $P_{30-24}$ |
| **QP 25** | 0 | 0 | 0 | 0 | 0 | 0 | 0.05 | 0 | 0 | 0.01 | 0.09 | 0.18 |
| **QP 30** | 0.43 | 0 | 0 | 0.48 | 0 | 0 | 0.27 | 0 | 0 | 0.33 | 0.38 | 0.49 |
| **QP 35** | 0.41 | 0 | 0 | 0.61 | 0.02 | 0.07 | 0.53 | 0.05 | 0.28 | 0.59 | 0.41 | 0.51 |
| **QP 40** | 0.63 | 0 | 0 | 0.19 | 0.14 | 0.24 | 0.64 | 0.30 | 0 | 0.60 | 0.67 | 0.75 |

*Remark*: We also compared the Rate Distortion (RD) of the test sequences at different frame rates. Fig. 8 illustrates the RD curves for different test sequences, where the distortion is measured based on PSNR. Although PSNR is widely used as a measure for distortion in compression, it is widely

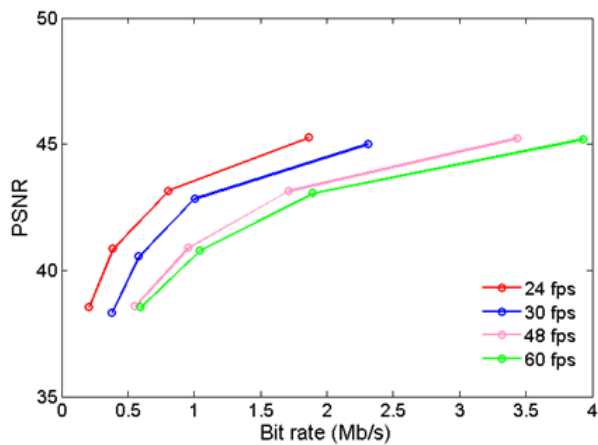
a) High motion: "Running" sequence

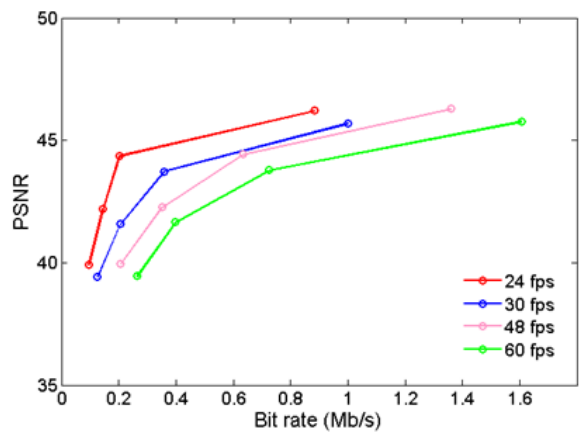
b) Medium motion: "BallSpin" sequence

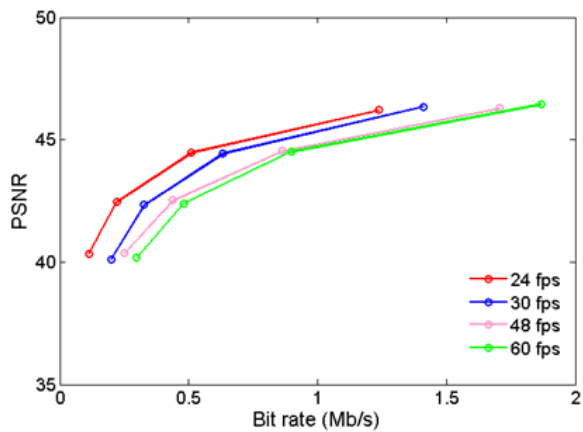
c) Low motion: "Forehand" sequence

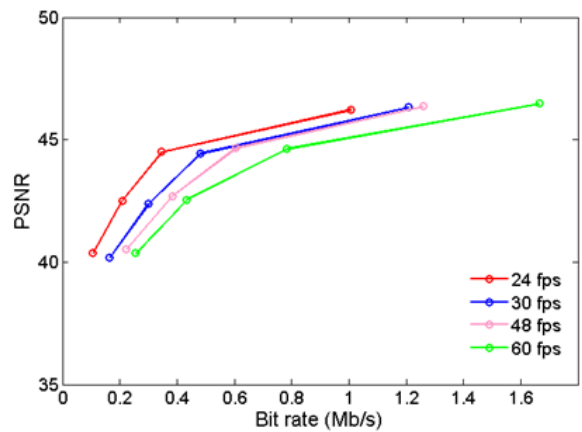
d) Very low motion: "Studying" sequence

**Fig. 8.** Rate-Distortion curves: PSNR at different bitrates for various frame rates, in four different levels of motion: a) High motion, b) Medium motion, c) Low motion, and d) Very low motion

accepted that it does not show high correlation with the perceptual quality of 3D content [36-40]. For instance, as it is observed from Fig. 8, the quality of the 24fps version of each test video is higher than other frame-rate versions of the same test sequence at all the bit rate levels. These results do not align with the subjective test results (see MOS values in Fig. 7). The main reason is due to the fact that PSNR does not take into account the temporal aspects of the video quality. Among these temporal video quality aspects, the worst-section-quality-effect and recency effect are two of the most important factors in video quality evaluation [41,42]. Presence of a section with poor quality, even for a very short period, in the video will highly change the overall visual quality of the video sequence. In addition, viewers tend to remember the quality of the most recent period of watched video (Like a hysteresis type effect) [41,42].

The reported PSNR values in Fig. 8 are calculated by finding the PSNR for the luma component of each frame, and then calculating the average value over all the frames. The overall video quality is affected by the quality of individual video frames as well as the temporal effects in the video [41]. In other words, the overall subjective quality of a video is not equal to the average quality of the frames, and it is for that reason that objective video quality metrics include temporal pooling techniques [40-42]. Extensive research has been done for designing temporal pooling strategies that combine the quality of individual frames to a single overall video quality score [41]. The reported results in Fig. 8 suggest designing an effective objective metric for 3D video applications that can be utilized in rate allocation and transmission of 3D video content, when different frame-rate versions of the video are available.

## 5. Conclusion and Future Work

In this paper, the relationship between the 3D quality of experience, bitrate, and frame rate was explored. First, a database of 3D sequences was created, involving scenes with different motion levels and frame rates of 24 fps, 30 fps, 48 fps, and 60 fps. Then, the quality of these videos was subjectively evaluated. Results of this experiment showed that subjects clearly prefer 3D videos with higher frame rates (48 and 60 fps) as there is a significant improvement in 3D quality when higher frame rates are used. Moreover, the same experiment revealed that increasing the frame rate to more than 60 fps, does not noticeably improve the 3D video quality.

In the second experiment, the stereoscopic scenes with four different frame rates were encoded at four compression levels (QPs of 25, 30, 35, and 40). Subjective quality evaluations of these 3D videos showed that for scenes with fast moving objects, the effect of frame rate on the overall perceived 3D quality is more dominant than the compression effect, whereas for scenes with low motion levels the frame rate does not have a significant impact on the 3D quality. In other words, higher frame rates improve the 3D QoE significantly when there is fast motion in a scene. In addition, high frame rate 3D videos with higher compression rates are preferred over slightly compressed but low frame rate 3D videos. The subjective test results suggest that in cases where the available bandwidth for video transmission drops (variable bandwidth channel), reducing the frame rate instead of increasing the compression ratio helps achieve the maximum possible 3D quality of experience level with respect to bandwidth.

In summary, our study suggests that the best practical frame rate for 3D video capturing is 60 fps, as it delivers excellent quality of experience and producing such content is possible by using available capturing devices. In fact, going beyond this frame rate does not yield visually noticeable improvement while the required effort and resources are not justifiable.

Future work will involve development of a frame-dropping based rate-adaptation scheme for 3D video content transmission, so that when the available bandwidth drops, the video frames that have low impact on the 3D quality of experience are selected and discarded before transmission. In particular, factors such as depth, motion, and also 3D visual attention models will be considered in selecting such frames. This rate-adaptation approach will allow adjusting the bitrate of the content with respect to the available bandwidth, while delivering the maximum possible 3D quality of experience.

## Acknowledgement

This work was partly supported by Natural Sciences and Engineering Research Council of Canada (NSERC) and the Institute for Computing Information and Cognitive Systems (ICICS) at UBC.